\documentstyle[12pt,blois]{article}
\def\Lsun{\ifmmode {\rm\,L_\odot}\else ${\rm\,L_\odot}$\fi}
\def\Msun{\ifmmode {\rm\,M_\odot}\else ${\rm\,M_\odot}$\fi}

\def\plotfiddle#1#2#3#4#5#6#7{\centering \leavevmode
\vbox to#2{\rule{0pt}{#2}}
\includegraphics{#1}}
\begin{document}
\heading{The Fraction of High Redshift Galaxies \\
in Deep Infrared Selected Field Galaxy Samples}

\author{P. Eisenhardt$^{1}$, R. Elston$^{2}$, S.A. Stanford$^{3}$, M. Dickinson$^{4}$, 
H. Spinrad$^{5}$, D. Stern$^{5}$, A. Dey$^{6}$} 
{$^{1}$ Jet Propulsion Laboratory, California Institute of Technology, Pasadena, USA.}  
{$^{2}$ University of Florida, Gainesville, USA.}
{$^{3}$ Lawrence Livermore National Laboratory, Livermore, USA.}
{$^{4}$ Johns Hopkins University, Baltimore, USA.}
{$^{5}$ University of California, Berkeley, USA.}
{$^{6}$ National Optical Astronomy Observatories, Tucson, USA.}

\begin{bloisabstract}

Recent results on the incidence of red galaxies in a $>100$
square arcminute field galaxy survey to $K=20$ and a $K=22$ survey of
the Hubble Deep Field are presented. We argue that a simple photometric
redshift indicator, based on $J-K$ color and supported by spectroscopic
results obtained with Keck, gives a reliable lower limit  of $\sim 25\%$ 
for the fraction of $z > 1$ galaxies in the 100 square arcminute
survey.  This fraction is substantially higher than found in previous
smaller samples, and is at least as consistent with predictions for
pure luminosity evolution as with those for hierarchical models.  The
same technique yields a very low fraction for the HDF, which appears to
be unusually underabundant in red galaxies.

\end{bloisabstract}

\section{Introduction}
Near-infrared luminosity provides a good measure of a galaxy's mass,
over a wide range of Hubble types \cite{Gavazzi}, redshifts, and
star formation histories \cite{Charlot},\cite{KC}.  With the availability
of sensitive, large-format infrared array cameras on large telescopes, it
is now practical to obtain infrared galaxy samples reaching below $L^*$ at
$z > 1$ over areas large enough to encompass hundreds of such galaxies.
In late 2001, the Space Infrared Telescope Facility (SIRTF \cite{Fanson}) 
will provide $\mu$Jy-level sensitivity in the mid-infrared, enabling 
rest-frame $2\ \mu$m-selected samples reaching $L^*$ at $z>3$ to 
be obtained \cite{Fazio}.
      
Kauffmann \& Charlot\cite{KC} have recently proposed that the fraction
of $z > 1$ galaxies in deep infrared $K$-selected samples provides a
powerful means of discriminating between pure luminosity evolution
(PLE) and hierarchical (CDM) scenarios for massive galaxy formation and
evolution.  They argue that the paucity of $z > 1$ galaxies in the
Hawaii $K$-band samples \cite{Songaila}, \cite{Cowie} already provides
strong evidence against the PLE scenario and is consistent with
$\Omega=1$ CDM models.  Similar arguments have been made based on the
absence of red objects in surveys covering several square arcminutes to
$K \sim 22$ \cite{Zepf}, primarily the KPNO 4m Infrared Imager
observations of the Hubble Deep Field (HDF IRIM) obtained by Dickinson
et al.

The HDF spans a volume small enough that it would be expected to
contain only a few dozen L* galaxies
with $1 < z < 2$ \cite{Dickinson}.  In combination with the strong
clustering seen in Lyman break galaxies over substantially 
larger fields \cite{Adelberger}, this suggests that global
conclusions drawn from samples like the HDF should be treated with
some caution.  

%\section{The EES Survey}

Using the KPNO 4m, Elston, Eisenhardt, and Stanford (hereafter EES)
have recently completed a substantially larger $K$-selected survey, 
whose properties are summarized and compared to the Hawaii and HDF IRIM
surveys in Table 1.  The EES survey is divided into 4 regions around
the sky, providing some indication of field-to-field variations caused
by clustering. Figure 1 shows a color--color diagram for the survey.
 
\medskip
\begin{center}
{\bf Table 1.} IR Field Survey Characteristics.
\end{center}
\begin{center}
\begin{tabular}{|c|c|c|c|c|c|}
\hline 
Survey & Bands & $K(10\sigma)$ & Area & N$_{gal}$ & N/A \\
       &       &  mag & sq arcmin &  & \#/sq arcmin \\
\hline
\hline
Hawaii\cite{Cowie}    & $B,I,K$ & 19.3 & 26 & 122 & 5 \\
\hline
EES & $B,R,I,Z,J,K$ & 20 & 124 & 1683 & 14 \\
    &               & 19 &     & 720  & 6 \\
\hline
HDF-IRIM & $J,H,K$ & 21.2 & 7 & 149 & 21 \\
         &        & 20   &   & 76  & 11 \\
         &        & 19   &   & 44  & 6 \\
\hline
\end{tabular}
\end{center}
%\medskip

%\clearpage
\begin{figure}
%\plotfiddle{rjk.eps}{3.0truein}{0}{70}{70}{30}{-180}
\plotfiddle{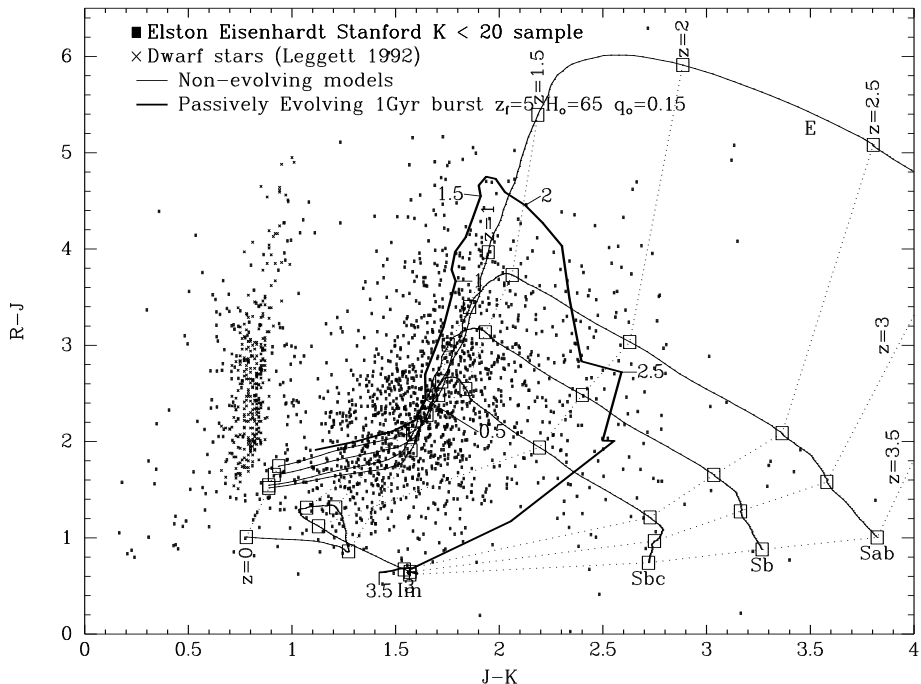}{3.0truein}{270}{50}{50}{-180}{270}

\caption{The $R-J$, $J-K$ color-color diagram for the EES survey.  
Overlaid on the data are the tracks of SEDs associated with zero
redshift galaxies of the indicated types.  Iso-redshift
contours are plotted at redshift intervals of 0.5 from redshifts of 0
to 3.5.    Also plotted on the
figure are the observed $RJK$ colors of dwarf stars from Leggett (1992). }

\end{figure}

\section{Extremely Red Objects}

The surface density of "extremely red objects" (ERO's) in the EES
sample is of interest.  Graham \& Dey \cite{GD} and Cimatti
\cite{Cimatti} define ERO's as objects with $R-K > 6$ and show that at
least one such source (Hu \& Ridgway 10) is a dusty galaxy at $z=1.44$
with detectable submm emission, implying L$_{bol} > 10^{12}$\Lsun and a
star formation rate of several hundred \Msun per year.  If the recently
identified population of field sources with a surface density $\sim1$
per square arcmin at comparable sub--mm fluxes  \cite{Hughes} are
similar in nature, they would dominate the global star formation rate.
SIRTF will be able to characterize this population from 3.6 to
160 $\mu$m with relative ease.

We find 0.7 sources per square arcminute with $R-K > 6$ and $K < 20$.
At this meeting, Barger defined very red galaxies by $I-K> 4$, finding
16 such objects to $K=20.1$ in a 62 square arcmin survey centered on
the HDF (i.e. 0.26 per square arcmin).  We find an average of 2.5
sources per square arcmin (ranging from 2.1 to 3.4 among the four
EES regions) with $I-K > 4$ and $K < 20$.  The reason for this
discrepancy is uncertain, but we suspect that at least part of the
answer is that the HDF simply has an unusually low abundance of red
galaxies.

\section{$J-K$ as a Lower Limit to the $z>1$ fraction}

Next we consider a different measure of the red population: $J-K > 1.9$.
Figure 1 shows that $J-K$ is primarily sensitive to redshift, at least
for earlier type galaxies.  A galaxy with the spectral energy distribution
of a present day elliptical galaxy would have $J-K\approx1.9$ at $z=1$.
Passive or active evolution will tend to make colors bluer,
so the fraction of galaxies with $J-K>1.9$ is a reasonably reliable lower
limit to the fraction of galaxies with $z > 1$.  Figure 2 demonstrates that
this assertion holds up under spectrosscopic scrutiny: while there
are indeed galaxies with $J-K < 1.9$ and $z > 1$, only one object has
$J - K > 1.9$ and $z < 1$.  

\begin{figure}
%\plotfiddle{lynxjmkvsz.ps}{5.0truein}{270}{70}{70}{-30}{400}
\plotfiddle{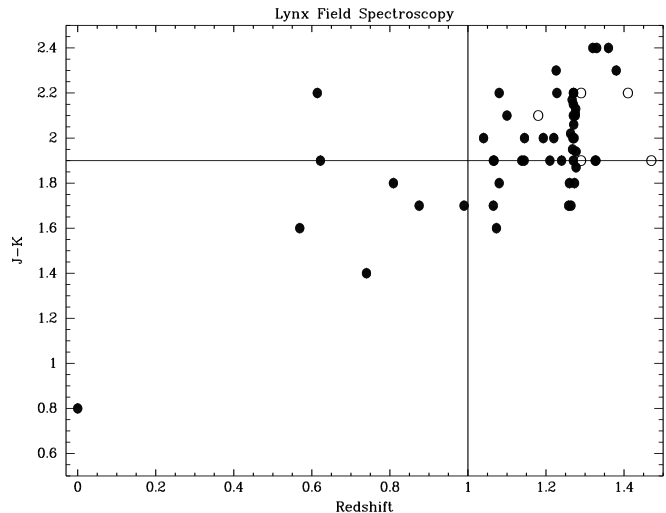}{2.3in}{270}{36}{36}{-140}{200}
%\plotfiddle{lynxjmkvsz.ps}{2.5truein}{270}{45}{45}{60}{240}
%\plotone{lynx_z_vs_jmk.ps}
\caption{$J-K$ vs. redshift for the objects in
the Lynx portion of the EES survey for which we 
have obtained slit mask spectra using the LRIS at Keck.}
\end{figure}

In Table 2 we list the lower limit to the fraction of galaxies with
$z>1$ calculated from the $J-K>1.9$ criterion for the EES and HDF IRIM
samples, together with lower and upper limits determined from
spectroscopy of the Hawaii and HDF IRIM samples, and predictions from
Kauffmann and Charlot \cite{KC}.  The spectropscopic lower limits
assume that none of the objects in the sample with unknown redshifts
lie at $z>1$, while the upper limits assume that all unknown redshifts
are at $z > 1$.

While the lower limits from the $J-K$ method for the EES sample are not
as high as the fractions predicted for the PLE scenario \cite{KC}, they
are consistent with PLE, and not particularly supportive of the
hierarchical model predictions.  Thus we consider the PLE scenario
still viable, at least based on the probable redshift distribution of
faint $K$--selected galaxy samples.  We have obtained hundreds more
spectra for the EES sample with LRIS and Keck in the fall of 1998, so
we expect to determine the redshift distribution of the sample with
higher confidence in the near future.  We also look forward to repeating
the test using SIRTF to obtain a rest-frame $K$-selected sample out to $z\sim 3$. 

\medskip
\begin{center}
{\bf Table 2.} Fraction of $z > 1$ Galaxies in IR Field Samples.
\end{center}
\begin{center}
\begin{tabular}{|c|c|c|c|c|c|c|}
\hline 
$K$ & \multicolumn{2}{c|}{K\&C 98} & Hawaii & EES & \multicolumn{2}{c|}{HDF IRIM} \\
(mag) & PLE & Hier & Spec & $J-K>1.9$ & Spec & $J-K>1.9$ \\
\hline
16--18 & 28\% & 0\% & 2--11\% & $>15\%$ & 14\% & $>0\%$ \\
18--19 & 54\% & 3\% & 10--17\% & $>23\%$ & 5--50\% & $>5\%$ \\
19--20 & & & & $>28\%$ & 15-52\% & $>9\%$ \\
20--21 & \raisebox{1.5ex}[0cm][0cm]{\}75\%} & \raisebox{1.5ex}[0cm][0cm]{\}20\%} & 
& & 17--70\% & $>3\%$ \\
21--21.5 & & & & & 12--88\% & $>17\%$ \\
\hline
\end{tabular}
\end{center}
%\medskip

\section{How Representative is the HDF?}

The lower limit to the fraction of $K$--selected $z>1$ galaxies in the
HDF determined via the $J-K > 1.9$ criterion is very low, and
reasonably consistent with the hierarchical model predictions of
\cite{KC} (although the spectroscopically determined limits for this
fraction are much less conclusive.) Red objects appear to be uncommon
in and around the HDF - a fact used by Zepf \cite{Zepf} and  Barger to
argue against a significant population of passively evolving elliptical
galaxies at high redshift.  This may be due to clustering effects,
since early type galaxies are strongly clustered at the present epoch.

We have examined the variation in surface density of $J-K>1.9$ and $K <
20$ objects within the EES survey.  Although the mean surface density
for the EES sample is 3.4 such objects per square arcmin, this value
ranges from 1.0 to 6.7 in 16 EES subfields the size of the HDF.  The
value in the HDF itself is 0.6, reinforcing the impression that the red
population HDF is unusually sparse.  Results from the HDF-South should
help settle this question.  The total surface density of the
HDF for all colors for $K < 20$ is also low, but within the EES range:
11.8 per square arcmin, whereas the EES subfields range from 11.8 (two
subfields) to 25.8.  This latter field contains a z=0.58 Rosat cluster,
and the next highest density is 16.9.

The alert reader may have noticed the clump of $z=1.27$ redshifts in
Figure 2.  This cluster in the Lynx EES field was identified by Stanford et al \cite{Stanford}.  Thus it is fair to ask how representative is the EES sample, 
or at least the Lynx portion of it (which also includes the $z=0.58$ cluster)?  
The surface density of $J-K>1.9$, $K<20$ objects in Lynx is 3.6 per
square arcmin, and 16 per square arcmin for all $J-K$, vs. 3.4 and 13.5
for these values respectively for the entire EES sample.  The EES sample
comoving volume out to $z\sim 2$ is a few $10^5$ Mpc$^3$.  If the present
number density of clusters ($\sim 10^{-5}$ Mpc$^{-3}$) does not evolve
rapidly, the presence of these clusters is not surprising.

\acknowledgements{The EES and HDF IRIM surveys would not have been
possible without generous allocations from the KPNO time allocation
committee.  We thank the organizers for putting together an extremely
stimulating and enjoyable meeting, capped by a truly spectacular
conference dinner in the Chateau de Chambord.  Portions of the research
described in this paper were carried out by the Jet Propulsion
Laboratory, California Institute of Technology, under a contract with
NASA.}

% References listed in alphabetical order ...

%
\vfill

\begin{bloisbib}
\bibitem{Adelberger} Adelberger, K. et al. 1998, \apj {505} {18}
\bibitem{Cimatti} Cimatti, A. et al. 1998, \nat {392} {895}
\bibitem{Charlot} Charlot, S. 1998, this volume
\bibitem{Cowie} Cowie, L., Songaila, A., Hu, E., \& Cohen, J. 1996, \aj {112} {839} 
\bibitem{Dickinson} Dickinson, M. 1998, in {\it Looking Deep in the Southern Sky},
Proceedings of the ESO/ATNF Workshop, eds. R. Morganti \& W. Couch, Springer, in press
\bibitem{Fanson} Fanson, J. et al. 1998, in 
{\it Space Telescopes and Instruments V} eds P. Bely and J. Breckinridge, 
Proc. SPIE 3356, 478
\bibitem{Fazio} Fazio, G., Hora, J., Stauffer, J., \& Eisenhardt, P. 1999,
in {\it Astrophysics with Infrared Surveys: A Prelude to SIRTF}, PASP Conference
series, eds M.D. Bicay et al, in press
\bibitem{Gavazzi} Gavazzi, G., Pierini, D., \& Boselli, A. 1996, \aa {312} {397}
\bibitem{GD} Graham, J.R. \& Dey, A. 1996, \apj {471} {720}
\bibitem{Hughes} Hughes, D.H. 1998, \nat {394} {241}
\bibitem{KC} Kauffmann, G., Charlot, S. 1998, \mnras {297} {L23}
\bibitem{Leggett} Leggett, S.K. 1992, \apjs {82} {351}
\bibitem{Stanford} Stanford, S.A. et al 1997, \aj {114} {2232}
\bibitem{Songaila} Songaila, A., Cowie, L., Hu, E., \& Gardner, J. 1994,
\apjs {94} {461}
\bibitem{Zepf} Zepf, S. 1997, \nat {390} {377}
%\bibitem{GG} Genius G., 1995, {\it preprint}
%\bibitem{jk} Kirk J., 1995, \aa {100} {44}
%\bibitem{wd} Mouse M., Duck D., 1935, \apj {1} {2}
%\bibitem{conf} Smith G., 1995, in {\it Proceedings of the nth Conference}
%     p. 129, eds Durand et al., World Editions
%\bibitem{} Lennon, J., McCartney, P., Harrison, G. \& Starr, R. 1962, \mnras
%{100} {1}
\end{bloisbib}
\end{document}